\documentclass[12pt,a4paper]{article}
\usepackage{jcappub}
\usepackage{bm}
\usepackage{indentfirst}
\usepackage{amsmath}
\usepackage{graphicx}
\usepackage{float}
\usepackage{amssymb}
\usepackage{subfigure}
\usepackage{hyperref}
\usepackage{array}
\usepackage{amsthm}
\usepackage{mathrsfs}
\usepackage{color}
\usepackage{comment}
\usepackage[normalem]{ulem}
\usepackage{url} 
\usepackage{orcidlink}
\hypersetup{
	colorlinks=true,
	linkcolor=red,
	citecolor=blue,
}
\allowdisplaybreaks[2]

\usepackage{xcolor}
\usepackage[normalem]{ulem}

\newcommand{\hws}{\textcolor{cyan}{$\mathcal{HT}$: }\bgroup\markoverwith{\textcolor{cyan}{\rule[.5ex]{2pt}{2.5pt}}}\ULon}

\def\apj{Astrophysical Journal}                

\begin{document}

\title{ Ensemble-Based Residual Tests of GW231123 across Waveform Models }

\author[a,1]{Dicong Liang,\note{Corresponding author.}}
\author[b,2]{Hai-Tian Wang,\note{Corresponding author.}}
\author[c]{Junlin Qin,}
\author[c]{Zhan-Feng Mai,}
\author[d]{Tong Jiang,}
\author[a]{Yingjie Yang}

\affiliation[a]{Department of Mathematics and Physics, School of Biomedical Engineering, Southern Medical University, Guangzhou, 510515, China}
\affiliation[b]{School of Physics, Dalian University of Technology, Dalian 116024, China}
\affiliation[c]{Guangxi Key Laboratory for Relativistic Astrophysics, School of Physical Science and Technology, Guangxi University, Nanning 530004, China}
\affiliation[d]{College of Physics and Technology, Kunming University, Kunming, Yunnan 650214, China}

\emailAdd{dcliang@smu.edu.cn}
\emailAdd{wanght9@dlut.edu.cn}
\emailAdd{junlin.qin@st.gxu.edu.cn}
\emailAdd{zf1102@gxu.edu.cn }
\emailAdd{jiangtong@hust.edu.cn}
\emailAdd{yyj@smu.edu.cn}

\abstract{
GW231123 is an exceptional gravitational wave event for which different waveform models yield significantly different inferred source parameters. 
Residual tests provide a direct way to assess whether each waveform model gives an adequate description of the observed signal. 
In this work, we extend the conventional residual-test methods by subtracting the 100 highest likelihood waveforms, rather than only the maximum likelihood waveform for each model, thereby propagating waveform reconstruction uncertainty into the residual analysis. 
This ensemble-based approach turns the residual test from a single waveform diagnostic into a robustness test over the local high likelihood waveform manifold. 
We further perform injection tests to quantify the detectability of cross-model waveform discrepancies in realistic detector noise. 
The large-scale implementation of these analyses is made possible by the high speed and low computational cost of our residual testing framework, which is based on three goodness-of-fit tests: the Kolmogorov-Smirnov test, the Anderson-Darling test, and Pearson's chi-squared test.
}


\maketitle

\section{Introduction}
\label{sec:intro}

On November 23rd 2023, at 13:54:30 UTC, a remarkable gravitational wave (GW) event, designated GW231123\_135430 (hereafter GW231123), was recorded by the two LIGO observatories \cite{LIGOScientific:2025rsn}. 
Bayesian inference using compact binary coalescence waveform models indicates that the event was produced by the merger of two massive black holes with a total mass $\gtrsim 190 M_\odot$ \cite{LIGOScientific:2025rsn}. 
The inferred component masses fall within the pair-instability mass gap predicted by standard stellar-evolution theory \cite{2019ApJ...887...53F,Woosley:2021xba}, posing a significant challenge to conventional black hole formation scenarios.
Various formation scenarios for the component black holes have been proposed and extensively studied, including direct core collapse \cite{Croon:2025gol,Gottlieb:2025ugy,Popa:2025dpz,Tanikawa:2025fxw}, hierarchical mergers \cite{Li:2025fnf,Stegmann:2025cja,Li:2025pyo,Paiella:2025qld,Passenger:2025acb,Liu:2025ogx,Liu:2025yok}, and primordial black-hole formation \cite{Yuan:2025avq,DeLuca:2025fln}.
Gravitational lensing offers another possible explanation for the large inferred masses, since lensing magnification can bias the observed masses upward relative to their intrinsic values. 
However, no evidence of lensing has been found for this event \cite{Chakraborty:2025pxt,Chan:2025kyu}.
While, it is pointed out in Ref.~\cite{Ray:2025rtt} that, microglitches can bias the measurements of black hole spins.
Another remarkable feature of GW231123 is the large inferred spins of its component black holes, which has motivated studies of accretion-driven spin evolution in Refs.~\cite{Bartos:2025pkv,Kiroglu:2025vqy}.
Beyond its astrophysical implications, GW231123 provides a valuable laboratory for testing fundamental physics.
The event has been employed to constrain axion models \cite{Caputo:2025oap}, search for deviations from the Kerr frequency spectrum \cite{Siegel:2025xgb},
test the wormhole echo hypothesis \cite{Lai:2026yvm,Angeloni:2026nmy},
investigate gravitational-wave polarization birefringence \cite{Liu:2026wor}, probe the no-hair theorem \cite{Wang:2025rvn}, among many other applications.

Five different waveform models for noneccentric, precessing binary black hole mergers are adopted to infer the properties of the source in Ref.~\cite{LIGOScientific:2025rsn}. 
Specifically, these waveform models are
{\tt NRSur7dq4}  \cite{Varma:2019csw},
{\tt SEOBNRv5PHM}  \cite{Ramos-Buades:2023ehm},
{\tt IMRPHENOMXPHM }  \cite{Colleoni:2024knd},
{\tt IMRPHENOMTPHM }  \cite{Estelles:2021gvs}, and
{\tt IMRPHENOMXO4a}  \cite{Thompson:2023ase}.
In general, numerical relativity (NR) simulations provide the most faithful description of the late inspiral, merger, and ringdown of compact binary coalescences, but they are too computationally expensive to be used directly in data analysis.
The {\tt NRSurrogate} models are constructed as reduced order interpolation models from a finite set of NR waveforms \cite{Blackman:2015pia,Blackman:2017dfb,Blackman:2017pcm,Varma:2019csw}. 
Thus, they can achieve high accuracy within the region of parameter space covered by the available NR simulations, but their validity is limited outside that region and for waveform durations beyond those in the training data. 
The {\tt SEOBNR} and {\tt PHENOM} model families are constructed by combining the analytic or semi-analytic waveform for the early inspiral stage and the late inspiral, merger and ringdown waveform from NR. 
As a result, these two model families have broader coverage on parameter space and can generate longer waveforms, although their accuracy depends on the approximations they used.
The {\tt SEOBNR} models are dynamical models based on effective-one-body (EOB) formalism, with waveform ingredients calibrated to NR information, particularly in the merger-ringdown regime \cite{Pan:2013rra,Purrer:2014fza,Purrer:2015tud,Babak:2016tgq,Bohe:2016gbl,Cotesta:2018fcv,Ramos-Buades:2023ehm}.
While the {\tt PHENOM} models are piecewise closed-form phenomenological
models, which uses physically motivated ansatze and fitted functional forms calibrated to EOB and NR information \cite{Ajith:2007qp,Ajith:2007kx, Hannam:2013oca,Schmidt:2014iyl,Khan:2015jqa,Khan:2018fmp,Khan:2019kot,
Estelles:2020osj,Estelles:2020twz,  
Hamilton:2021pkf,Ghosh:2023mhc,
Colleoni:2024knd,
Estelles:2021gvs,
Thompson:2023ase}.
The three {\tt PHENOM} models considered here, {\tt IMRPHENOMXPHM },  {\tt IMRPHENOMTPHM },  and {\tt IMRPHENOMXO4a}, improve the treatment of spin precession in different ways, relative to earlier {\tt PHENOM} models.

The event GW231123 shows exceptional properties that the spins of the component black hole are larger than 0.8 \cite{LIGOScientific:2025rsn}. 
It is challenging for all the five waveform models, since none of them is calibrated to NR waveforms from precessing binaries with such high spin. 
The waveform uncertainties might be the reason of the significant measurement differences of this event across different waveform models.
According to the posterior samples, multiple parameters fail to agree within $90\%$ credible interval \cite{LIGOScientific:2025rsn}. 
There are also other discussions about if the measurement differences originate from signal overlapping \cite{Hu:2025lhv}, or orbital eccentricity \cite{Jan:2025zcm}.

Since no single waveform model is clearly preferred, it is important to perform residual tests for all the five waveform models.
Residual tests are widely used in GW community to assess whether a waveform model provides an adequate description of the observed signal \cite{LIGOScientific:2016lio,LIGOScientific:2019fpa,LIGOScientific:2020tif,LIGOScientific:2021sio,LIGOScientific:2026qni,Green:2017voq,Nielsen:2018bhc,Marcoccia:2020rag,Liang:2020rt,Liang:2025zws}.
In Ref.~\cite{LIGOScientific:2025rsn}, residual data were constructed by subtracting the maximum likelihood waveform from the {\tt NRSur7dq4} parameter estimation samples, and then {\tt BAYESWAVE} \cite{Cornish:2014kda,Littenberg:2014oda,Cornish:2020dwh} was used to search for coherent power in the residuals.
A similar residual analysis based on {\tt IMRPHENOMXPHM} model was perform in Ref.~\cite{LIGOScientific:2026qni}.

In this work, we apply three different frequentist residual tests to GW231123, following the methodology of Ref.~\cite{Liang:2025zws}, using five waveform models.
In addition to the conventional maximum likelihood subtraction, we construct residuals using the 100 highest likelihood waveforms for each model, which allows us to test the robustness of the residual analysis against waveform reconstruction uncertainty. 
We further examine the waveform discrepancies between different models and use injection simulations to quantify whether such discrepancies are detectable in realistic detector noise.

The paper is organized as follows.
In Section \ref{sec:II}, we introduce the residual test methods used in this work.
Next, we characterize the 100 highest likelihood waveforms for the five waveform models in Section~\ref{subsec:waveforms} and apply the residual tests to them in \ref{subsec:residual}.
Then, we compare the waveform reconstructions across different models and investigate the detectability of their differences using injection simulations in Section~\ref{subsec:injection}.
Finally, in Section~\ref{sec:disc}, we summarize our results and discuss their implications.

\section{Preliminaries}
\label{sec:II}

Residual tests provide a broad check for whether a waveform model gives a satisfactory description of the observed strain data. 
If the waveform template, $h(t)$, captures the GW signal $s(t)$ accurately, then the signal-template difference $s(t)-h(t)$ should be sufficiently small at the noise level.
In this case, subtracting the template from the data, $d(t)$, should leave a residual, $r(t)$, that is statistically consistent with detector noise. 
Conversely, any significant non-noise-like structure in the residual may indicate waveform mismatch, unmodeled physical effects, or imperfect noise characterization. 
Therefore, residual tests are useful for assessing the strain-level consistency between the observed signal and a reconstructed waveform. 

In the conventional residual test in the literature \cite{LIGOScientific:2016lio,LIGOScientific:2019fpa,LIGOScientific:2020tif,LIGOScientific:2021sio,LIGOScientific:2026qni,Green:2017voq,Nielsen:2018bhc,Marcoccia:2020rag,Liang:2020rt,Liang:2025zws}, the waveform template to be subtracted is the so-called best-fit waveform, which is produced by the parameters with maximum likelihood from the inferred posterior samples. 
While in this work, we extend the residual analysis by using not only a single maximum likelihood waveform but also the 100 posterior samples with the largest likelihood values.
That is to say, we construct a set of residuals for each waveform model as follows,
\begin{align}
    r^{\rm M}_i(t) = d(t) - h^{\rm M}_i(t), \quad i=1,2,3,\cdot\cdot\cdot,100.
\end{align}
Here the superscript ``M" labels the five different models.
The subscript ``$i$" denotes the likelihood rank of the posterior samples, with $i=1$ corresponding to the maximum likelihood sample.
Therefore, for each waveform model and each detector, we obtain 100 residual timeseries rather than a single residual.

For a special event such as GW231123, whose source parameters lie in an extreme high-mass and high-spin region, the likelihood surface can be broad, irregular, and potentially multi-modal. 
The numerically identified ''maximum likelihood" sample is therefore not necessarily unique, nor should it be regarded as a physically privileged waveform. 
In practice, the reported maximum likelihood waveform is simply the loudest sample found by the sampler within a finite posterior set. 
Consequently, a residual test based on only this single best-fit waveform may be sensitive to sampling fluctuations or to one particular realization of the best-fit parameters. 
By instead using an ensemble of high-likelihood waveforms, we obtain a more robust assessment of the residuals and reduce the dependence of the test on any single posterior sample.

For these residuals, $r^{\rm M}_i(t)$, we apply the q-transform to them after whitening them.
For the whitening process, we use the 4096-second segment of the data to evaluate the power spectral density (PSD) of the noise with Welch method.
Specifically, we use the {\tt q-gram} module in {\tt GWpy} \cite{Macleod:2021goi} to calculate the q-transformed energies.
As a modification of the standard Fourier transform, the q-transform is widely used to visualize GW signals and map excess power in the time-frequency plane of noisy data \cite{Chatterji:2004qg,chatterji2005search,Blackburn:2008ah,Vazsonyi:2022jul}.
After normalization, the q-transformed energies of white noise are expected to follow the exponential distribution \cite{chatterji2005search,Vazsonyi:2022jul}.
Thus, our residual tests aim to check the consistency between the q-transformed energies of residuals and the expected exponential distribution.
In practice, we select only the energies computed
from one-segment data, which begins at 0.4 second before the merger time and ends at 0.1 second after. The frequency range spans from 20 Hz to 256 Hz, which contains most of the signal power. 
Then, we follow \citet{Liang:2025zws} to apply the three goodness of fit tests with these normalized q-transformed energies:
Kolmogorov-Smirnov test (KS test),
Anderson-Darling test (AD test),
and Pearson's chi-squared test ($\chi^2$ test).

The KS statistic is defined as the maximum absolute difference between the empirical cumulative distribution function of the samples $F_N(y)$, and the expected cumulative distribution function, $F(y)=1-\text{exp}(-y)$, evaluated over all values of $y$ \cite{an1933sulla,smirnov1948table}, that is,
\begin{align}
    \mathcal{S}^{\rm KS} = \sup_y|F_N(y)-F(y)|.
\end{align}
The Anderson–Darling (AD) statistic is similar to the KS statistic, but it gives more weight to the tails of the distribution. 
It is defined as \cite{anderson1954test}
\begin{align}
    \mathcal{S}^{\rm AD} = -N - \sum_{k=1}^N  \frac{2k-1}{N} 
    \left[ \ln(F(Y_k))-\ln(1- F(Y_{N+1-k}))  \right], 
\end{align}
where $N$ is the count of the data, and $Y_i$ denotes the data sorted in ascending order.
The $\chi^2$ statistic is the measurement of discrepancy between observed and expected frequencies in binned data \cite{pearson1900x}
\begin{align}
    \mathcal{S}^{\chi^2} = \sum^{n_b}_{j=1}\frac{[O_j-E_j]^2}{E_j},
\end{align}
where $n_b$ is the number of bins, $O_j$ and $E_j$ are the observed count and the expected count according to the exponential distribution in the $j$th bin, respectively.

\section{Residual Analysis}
\label{sec:III}

\subsection{Highest Likelihood Waveforms}
\label{subsec:waveforms}

The event GW231123 was recorded by the two LIGO detectors, i.e.,
LIGO Hanford (H1) and LIGO Livingston (L1), while Virgo and KAGRA
detectors were not online at that time.
The strain data and the posterior samples of parameter estimation are provided in the Gravitational Wave Open
Science Center \cite{Trovato:2019liz}  \footnote{ \url{https://gwosc.org/eventapi/html/O4_Discovery_Papers/GW231123_135430/v1/} }.

\begin{figure}
    \centering
    \includegraphics[width=\linewidth]{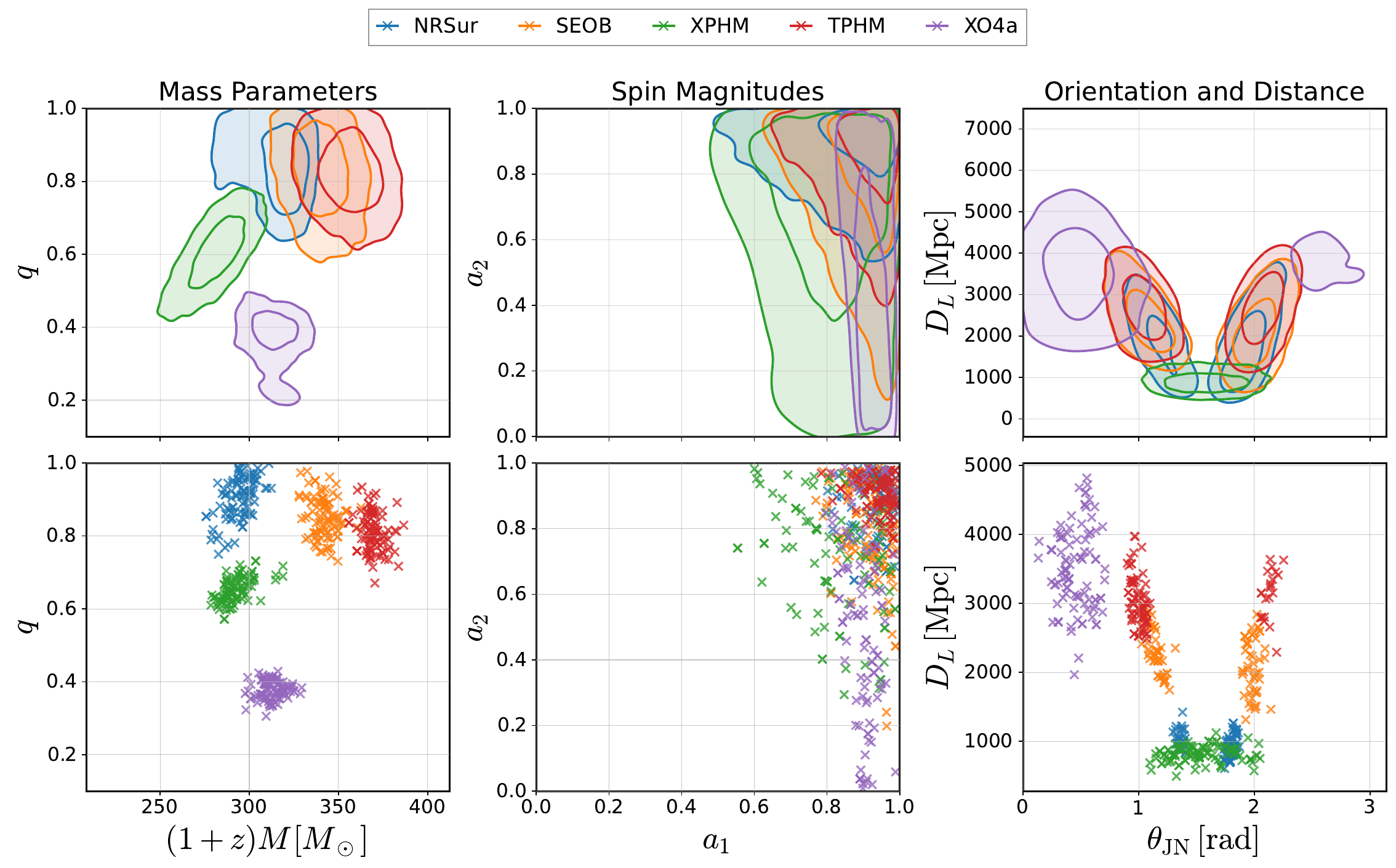} 
    \caption{Posterior distributions and high-likelihood samples for the five waveform models. 
    The upper panels show the posterior distributions of selected source parameters, with contours enclosing the 50\% and 90\% credible regions. 
    The lower panels show the corresponding 100 highest likelihood samples for each model. 
    From left to right, the columns display the redshifted total mass $(1+z)M$ and mass ratio $q$, two spin magnitudes $a_1$ and $a_2$, luminosity distance $D_L$ and orientation $\theta_{\rm JN}$.
    }
    \label{fig:para}
\end{figure}

We show the posterior distribution of six representative parameters, i.e., two mass parameters, two spin parameters and orientation and distance parameters, in the upper panels of Fig.~\ref{fig:para}.
For simplicity, the waveform models, {\tt  NRSur7dq4},
{\tt SEOBNRv5PHM},
{\tt IMRPHENOMXPHM},
{\tt IMRPHENOMTPHM}, and
{\tt IMRPHENOMXO4a}, are denoted as {\tt  NRSur}, {\tt SEOB}, {\tt XPHM}, {\tt TPHM}, and {\tt XO4a}, respectively
in this figure and hereafter.

As is shown in Fig.~\ref{fig:para}, for different models, the location of these parameters lies on significantly different region. 
In the fourth Gravitational-Wave Transient Catalog \cite{LIGOScientific:2025slb}, most events have consistent source parameters when inferring with different waveform models. 
In other words, the differences between results obtained with the different models are subdominant for most events, comparing to the statistical uncertainty induced by noise.
While, GW231123 shows largest systematics among all the exceptions (see Section 3.6 of Ref.~\cite{LIGOScientific:2025slb}).

\begin{figure}
    \centering
    \includegraphics[width=\linewidth]{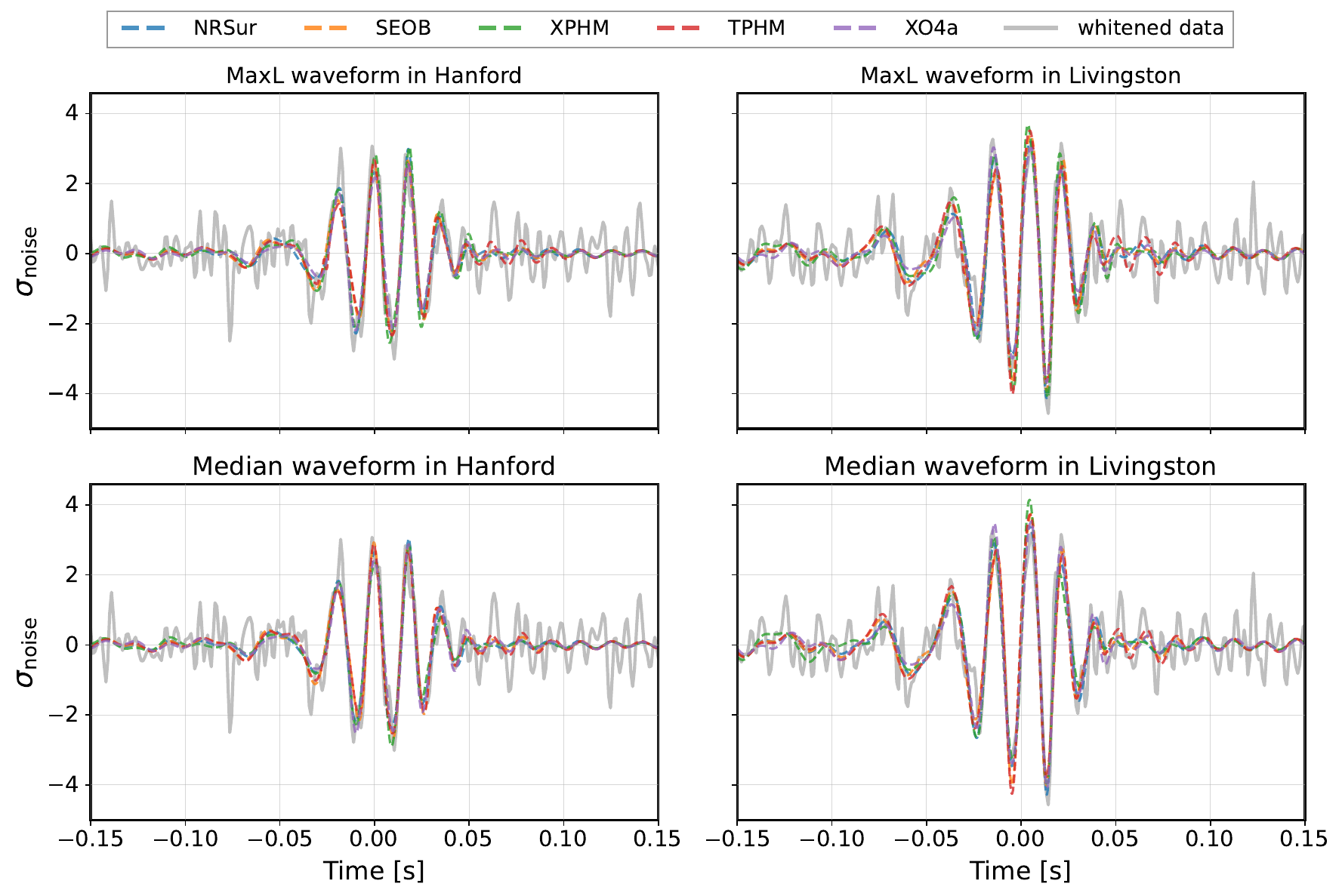} 
    \caption{Whitened waveform reconstructions of GW231123 in Hanford and Livingston. 
    The upper panels show the maximum likelihood waveforms from the five waveform models, while the lower panels show the median waveforms constructed from the 100 highest likelihood samples for each model. 
    The gray curves show the strain data, whitened and bandpass filtered between 20 and 256 Hz. 
    Time is measured relative to 13:54:30.619 UTC, November 23rd 2023. }
    \label{fig:waveforms}
\end{figure}

In the lower panels of Fig.~\ref{fig:para}, we show the parameters of the 100 posterior samples who has highest likelihood, which we will use to generate the waveform.
Although these parameters differ from each other within different models, the whiten time-domain waveforms they generate are very similar visually. 
We show the best-fit waveforms of the five models in the upper panels of Fig.~\ref{fig:waveforms}.
The waveforms are whitened, so they are in unit of the standard deviation with respect to the noise, i.e., $\sigma_{\rm noise}$.
For comparison, we add the whiten strain data in the figure. The strain data are further bandpassed in the frequency range from 20 to 256 Hz, so that to make the major cycles of the signal more visuable. 

In the lower panels of Fig.~\ref{fig:waveforms}, we show the median waveforms obtained from the 100 highest likelihood samples of each model. 
In both detectors, all models reproduce the dominant signal cycles around the merger, where the signal power is concentrated. 
The maximum likelihood waveforms are visually very similar across models, and the median waveforms in the lower panels closely follow the corresponding maximum likelihood reconstructions, demonstrating the internal stability of the high-likelihood waveform ensemble within each model. 
These results indicate that the strain-level reconstruction of the main GW231123 signal is robust, and that the large parameter-level waveform systematics do not immediately translate into large differences in the observed detector-frame waveform.

\begin{figure}
    \centering
    \includegraphics[width=\linewidth]{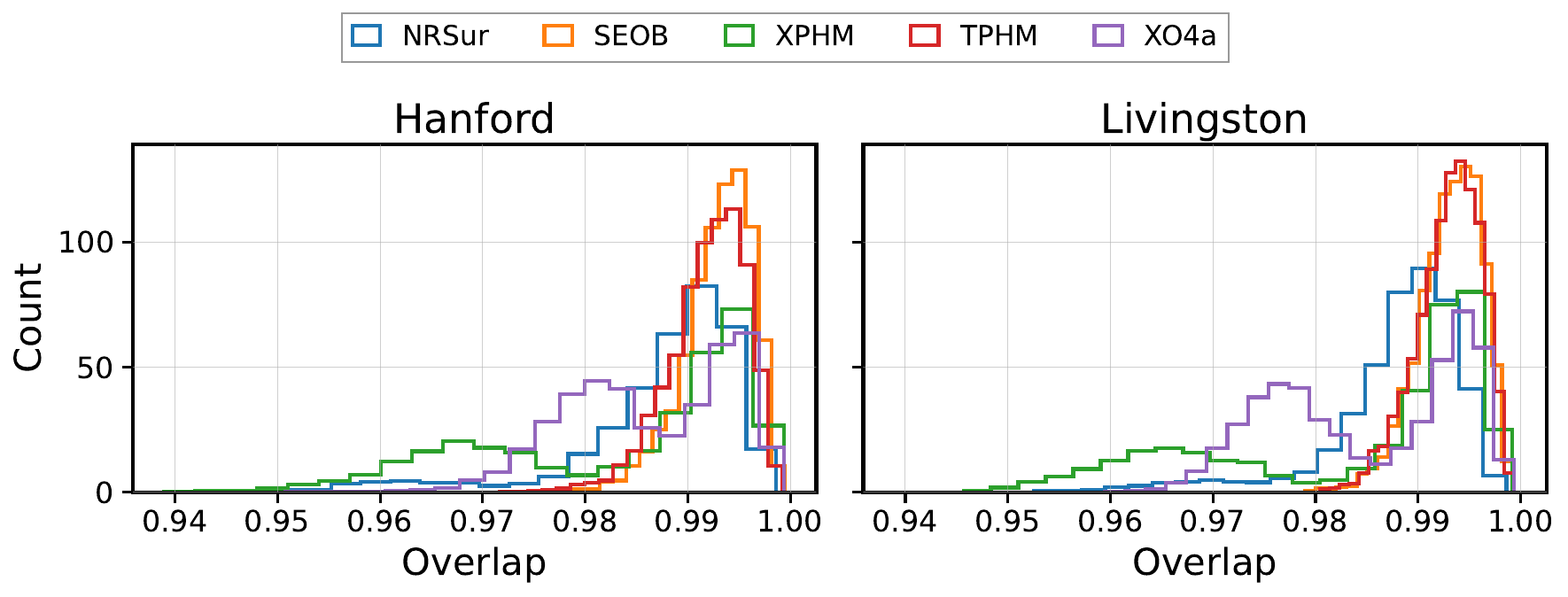} 
    \caption{Pairwise overlaps among the 100 highest likelihood waveforms for each waveform model in Hanford and Livingston.  }
    \label{fig:overlap}
\end{figure}

Statistically, we further calculate the overlap of the 100 highest likelihood waveforms.
The overlap of two waveforms is defined as
\begin{equation}
\mathcal{O}=
\frac{\langle h_1,h_2\rangle }
{\sqrt{ \langle h_1,h_1\rangle \langle h_2,h_2\rangle} }
\end{equation}
with the noise weighted inner product 
\begin{equation}
\langle h_1,h_2\rangle=
4\,\mathrm{Re} \int_{f_{\min}}^{f_{\max}}
\frac{\tilde{h}_1(f)\tilde{h}_2^*(f)}
{S_n(f)} \,df.
\end{equation}
Here $S_n$ is the PSD of the noise which we use to whiten the data. 
We adopt $f_{\rm min}=20\, \text{Hz}$ and $f_{\rm max}=256 \, \text{Hz}$.
A higher degree of similarity between two waveforms corresponds to an overlap value closer to $1$.
For each model, there are $C^2_{100} =4950$ pairs of waveforms and their overlap values are shown in Fig.~\ref{fig:overlap}.

Overall, the values of overlap are larger than 0.94 for all the five models, and the median values of overlap are all larger than 0.985 in both detectors.
This means that, although we use 100 high-likelihood samples, these waveforms are generally very similar to each other. 
Therefore, the ensemble residual test is not dominated by wildly different waveform subtractions.
The value of overlap has a similar distribution in both detectors.
The overlap of models {\tt SEOB} and {\tt TPHM} is highly concentrated in the range from 0.98 to 1.
While the overlap of models {\tt XPHM} and {\tt XO4a} has a wider spread.
The figure shows that the 100 selected waveforms are all near-optimal and mostly mutually consistent, but not identical. 
The variability of the waveform supports the use of an ensemble-based residual test rather than relying on a single maximum likelihood waveform.

\subsection{Residual Test}  
\label{subsec:residual}

\begin{figure}
    \centering
    \includegraphics[width=\linewidth]{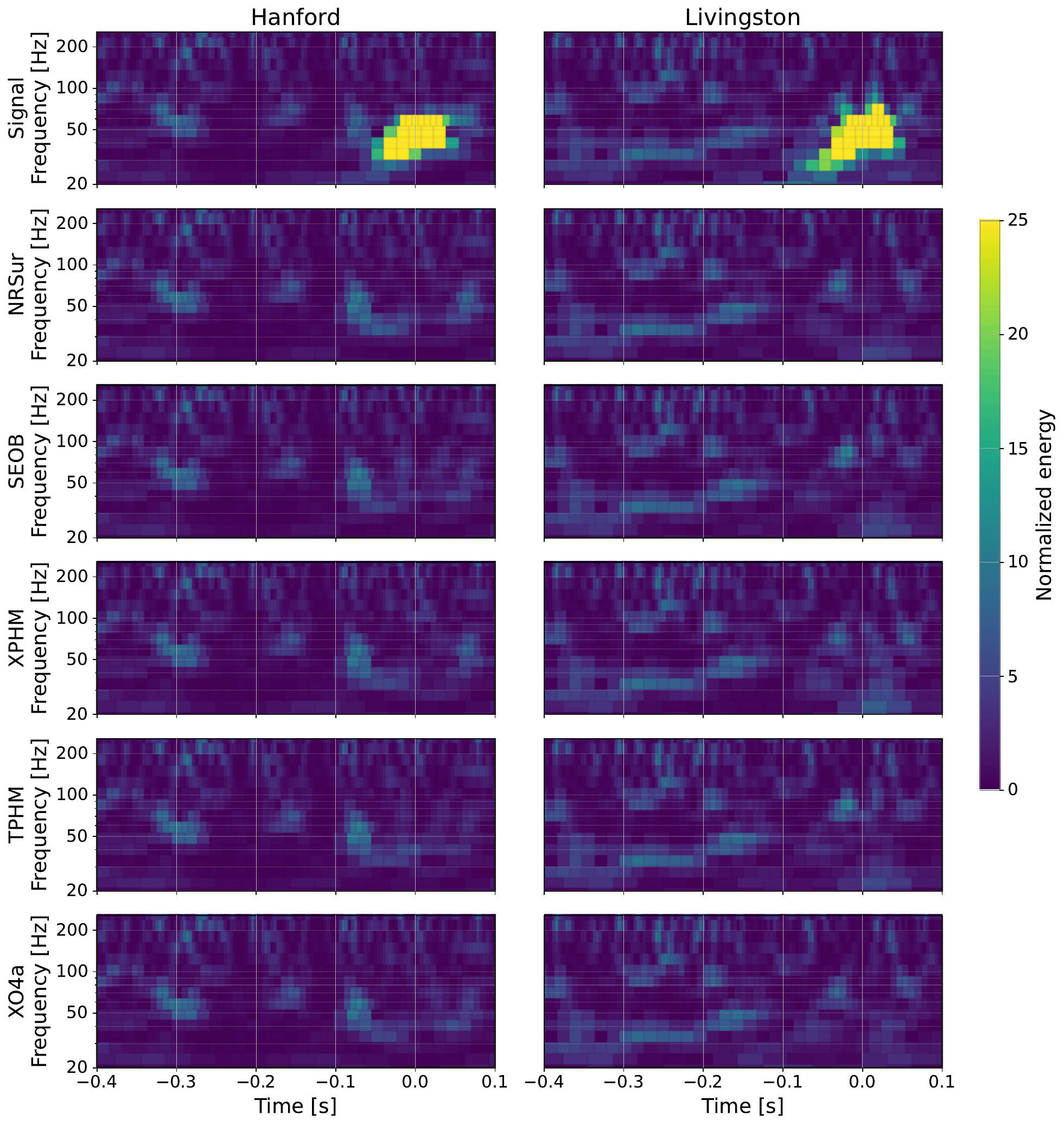} 
    \caption{ Normalized q-transformed energy maps for GW231123 in Hanford and Livingston.
    The top row shows the whitened strain data before subtraction.
    The lower rows show the median value of the normalized q-transformed energies of the residuals after subtracting the 100 largest likelihood waveforms for each waveform model.
    Time is measured relative to 13:54:30.619 UTC, November 23rd 2023.
    }
    \label{fig:q_median}
\end{figure}

After generating the 100 highest likelihood waveforms for each detector and for each waveform model, we subtract them from the strain data to get the residuals.
Then we apply q-transform to them and get the energy distribution.
The median values of these energies in the time-frequency plane are represented in Fig.~\ref{fig:q_median}.

For comparison, we show the energy distribution of the original strain in the first row in Fig.~\ref{fig:q_median}. 
The bright regions represent the short chirp-like excess power of the signals, which is clearly visible in both detectors.
In the residual panels, this coherent time-frequency structure is no longer present at a comparable level, indicating that the dominant signal power has been removed by all five waveform models. 
Any remaining discrepancy between the waveform templates and the observed signal is weak and appears noise-like at the level of the q-transform morphology.
Small model-dependent differences can still be seen in the residual structures, especially near the merger region where the signal power is strongest.

\begin{figure}
    \centering
    \includegraphics[width=\linewidth]{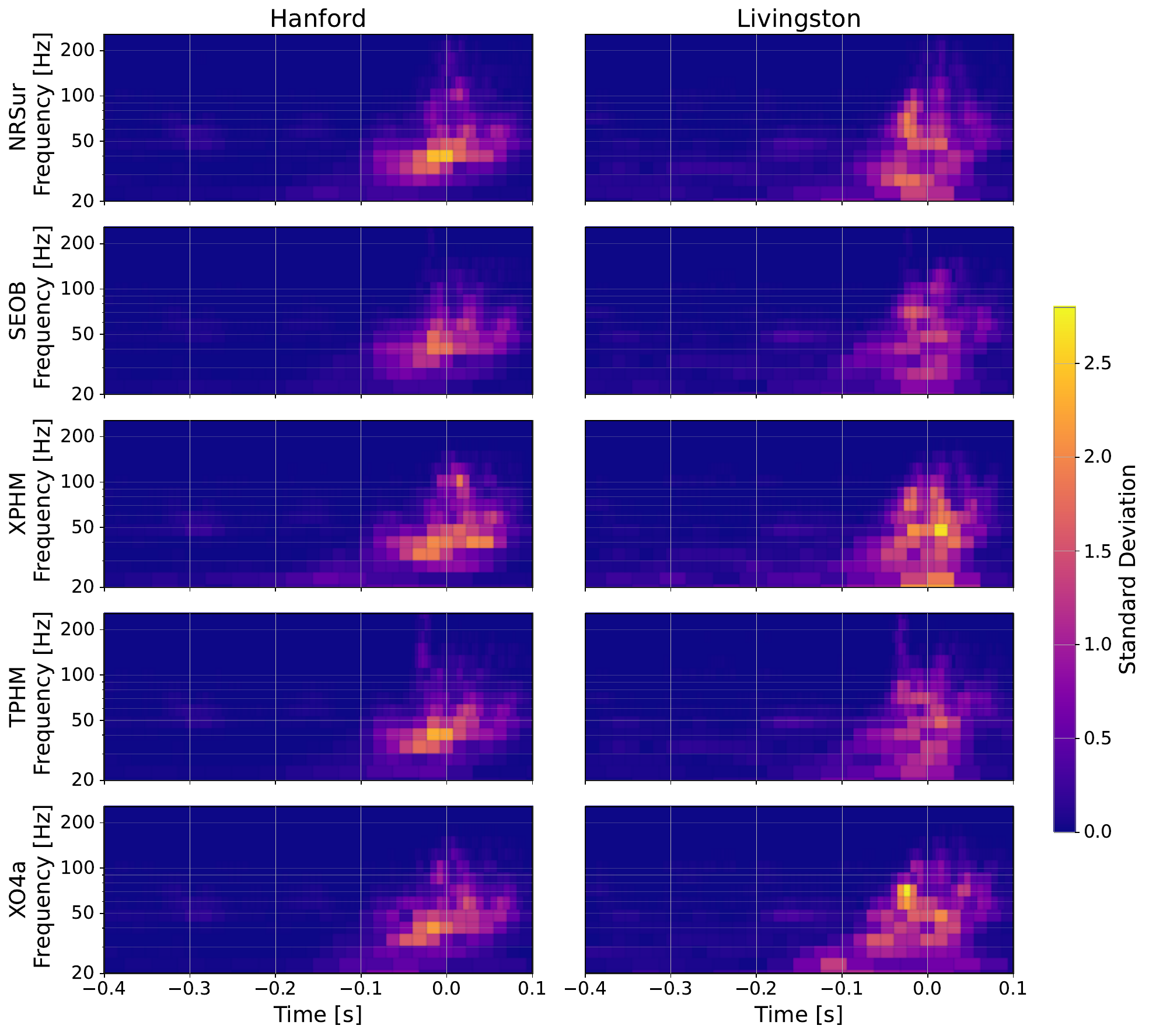} 
    \caption{Pixel-wise standard deviation of the normalized q-transform energies of the residuals for GW231123 in Hanford and Livingston. For each waveform model, we subtract the 100 highest likelihood waveforms from the data and compute the standard deviation of the resulting normalized q-transform energies at each time-frequency pixel. Bright regions mark where the residual energy varies most across the high-likelihood waveform ensemble.
    Time is measured relative to 13:54:30.619 UTC, November 23rd 2023.
    }
    \label{fig:q_std}
\end{figure}

The median alone does not reveal how much the residual maps vary within each model, so in Fig.~\ref{fig:q_std} we examine the pixel-wise standard deviation of the residual q-transform energies across the 100 highest likelihood waveform subtractions.
The internal waveform variation within each waveform model is presented in the time-frequency plane in the figure. 
For each model, it shows where the residuals differ from each other, and therefore directly measures the internal uncertainty of the waveform reconstruction. 
The variations are not uniformly distributed over the whole time-frequency map, but are instead concentrated around the merger and ringdown region. 
Comparing different models, the overall pattern of variation is similar, but the strength and extent of the high-variance regions differ, indicating that the internal stability of the high-likelihood waveform ensemble is model dependent.
The variation is more visible in Livingston due to its higher sensitivity.

\begin{figure}
    \centering
    \includegraphics[width=\linewidth]{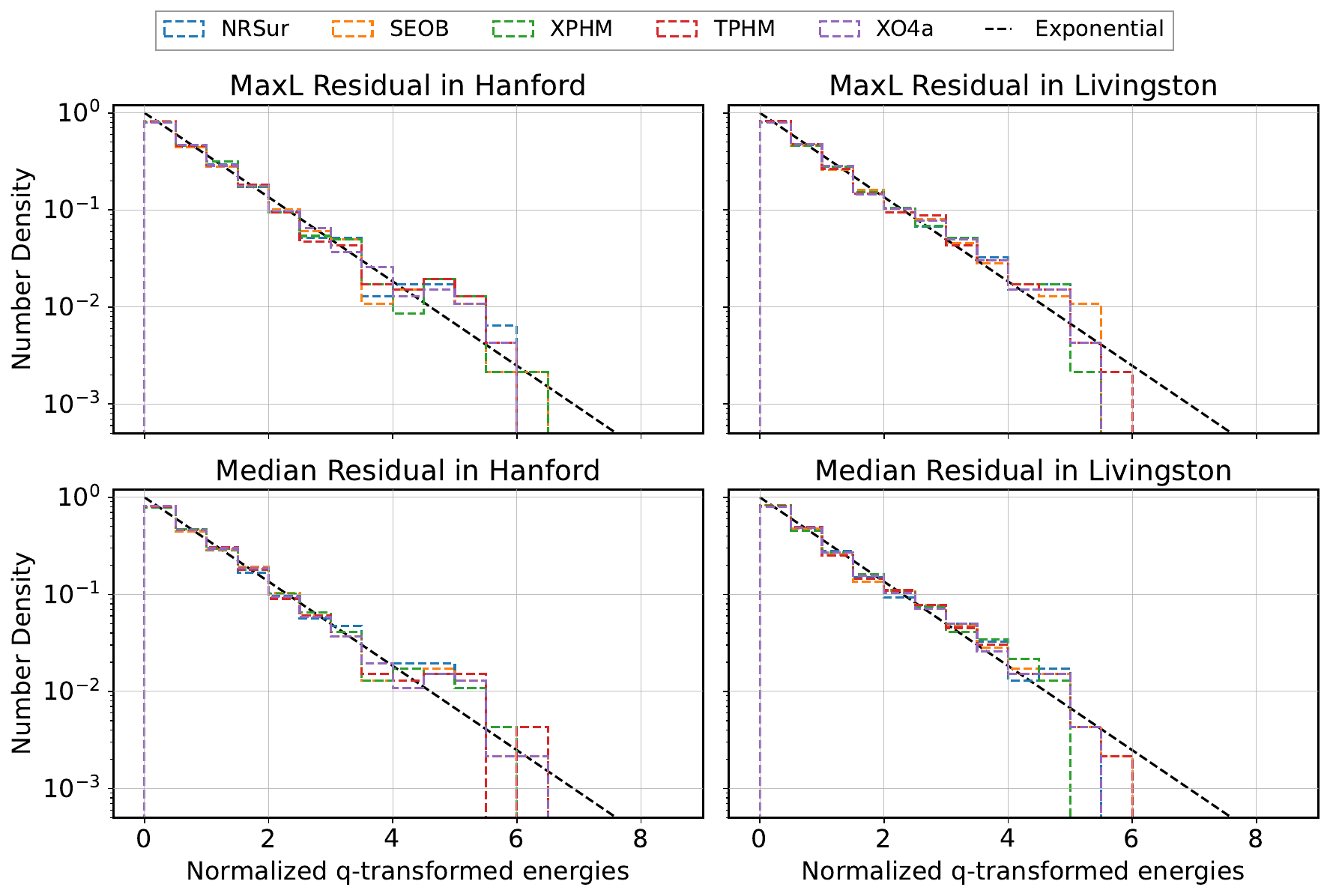} 
    \caption{Histograms of the normalized q-transform energies of the residuals in Hanford and Livingston. 
    The upper panels correspond to residuals obtained after subtracting the maximum likelihood waveform of each model. 
    The lower panels show the histograms computed from the median normalized q-transform energy maps of the residuals constructed from the 100 highest likelihood waveform subtractions. 
    The black dashed line shows the exponential distribution expected for white noise.
    }
    \label{fig:q_hist}
\end{figure}

After showing the characteristic of the residuals in the time-frequency plane in Fig.~\ref{fig:q_median} and Fig.~\ref{fig:q_std}, we next quantify whether these residuals are globally consistent with noise.
We normalize the q-transformed energies and present them into histograms in Fig.~\ref{fig:q_hist} and we show the expected exponential distribution (denoted by the black dashed lines) for comparison.

As is shown in Fig.~\ref{fig:q_hist}, the normalized q-transform energies of the residuals broadly follow the exponential distribution expected for stationary white noise. 
This agreement holds for both detectors and for all five waveform models.
Similarity can be found between the residuals produced from the maximum likelihood waveform and the median of the 100 high-likelihood residual maps. 
No model shows a clearly enhanced high-energy tail, suggesting that the dominant signal power has been removed and that the remaining residuals are globally consistent with noise.
The visual agreement motivates the quantitative goodness-of-fit tests presented below.

\begin{figure}[t] 
    \centering
    \includegraphics[width=\linewidth]{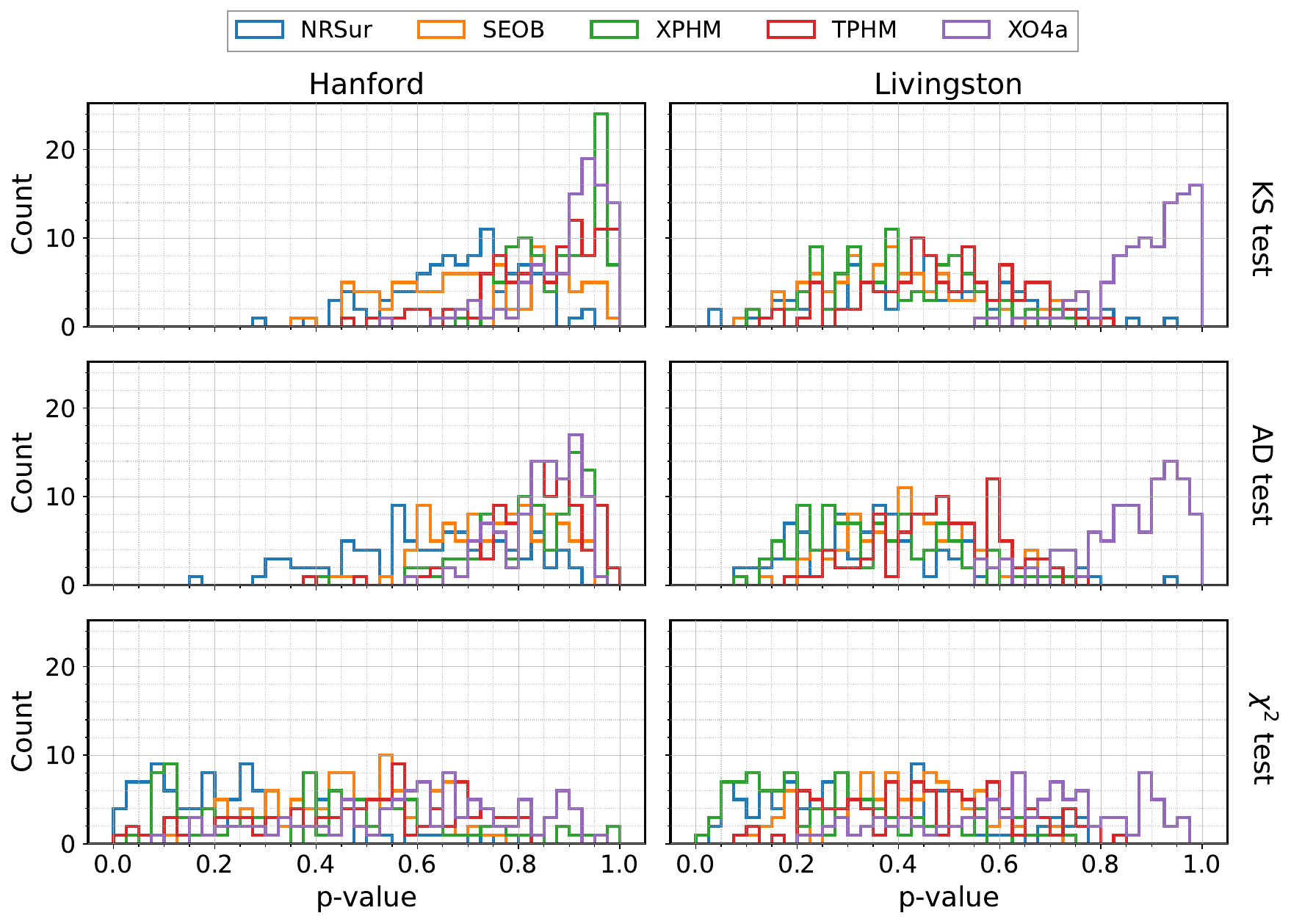} 
    \caption{Histograms of the p-values from the residual tests applied to the 100 residuals for each waveform model in Hanford and Livingston. 
    Each residual is obtained by subtracting one of the 100 highest likelihood waveforms from the strain data. 
    The three rows show the Kolmogorov–Smirnov, Anderson–Darling, and chi-squared tests. }
    \label{fig:p_res}
\end{figure}

We apply the three residual tests, namely the KS test, AD test, and $\chi^2$ test, to the 100 residuals for each detector and each waveform model. 
The resulting p-value distributions are shown in Fig.~\ref{fig:p_res}. 
These histograms quantify how the residual-test outcomes vary across the high-likelihood waveform ensemble. 
Most p-values are moderate or large rather than concentrated near zero, among which the smallest one is 0.004. 
It is indicated that the residuals are generally consistent with the expected noise distribution and only very few residuals show less consistency. 
The width of each histogram reflects the sensitivity of the residual-test result to the choice of high likelihood waveform within the same model. Differences among waveform models and between detectors show that the p-value distributions are affected by both waveform-reconstruction uncertainty and detector-specific noise fluctuations.

\begin{figure}[t] 
    \centering
    \includegraphics[width=\linewidth]{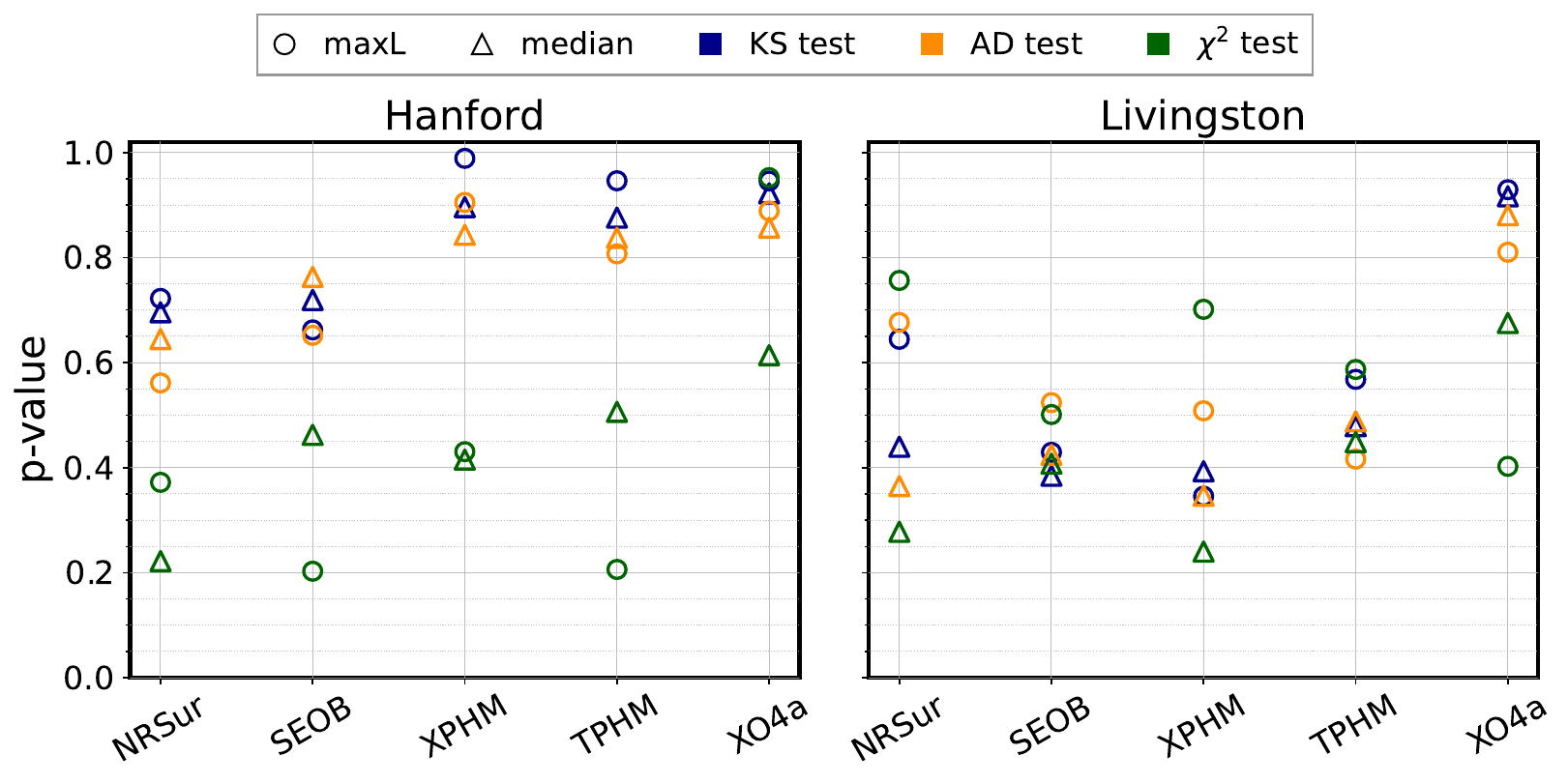} 
    \caption{ Summary of the residual test p-values for the maximum likelihood and median residuals in Hanford and Livingston. 
    For each waveform model, circles denote the p-values obtained from the residual after subtracting the maximum likelihood waveform, while triangles denote the median p-values among 100 high-likelihood residuals. 
	Specifically, the p-values for the best-fit waveform are denoted as `circle" and the median ones are denoted as ``triangle" in the figure.
     Colors indicate the three goodness-of-fit tests. }
    \label{fig:p_res_m}
\end{figure}

While Fig.~\ref{fig:p_res} shows the full distribution of p-values obtained from the 100 high-likelihood residuals, it is also useful to summarize these distributions with representative values. 
Therefore, we show the p-values obtained from the maximum likelihood residual and the median p-values among the high-likelihood residuals in Fig.~\ref{fig:p_res_m}.

For each waveform model and detector, the p-values of the three tests remain well above common significance thresholds, indicating no statistically significant deviation from the expected noise distribution. 
The maximum likelihood and median results are generally consistent, showing that the residual-test conclusion is not driven by a single posterior sample. 
Although the p-values vary among waveform models, detectors, and goodness-of-fit tests, no model produces systematically extremely small p-values. 
Therefore, the residuals of GW231123 are consistent with noise for all five waveform models considered here.
In other words, although the five waveform models exhibit noticeable waveform systematics, they still provide sufficiently accurate descriptions of the observed signal, comparing to the current detector sensitivity.

\subsection{Injection Test}
\label{subsec:injection}

A recent work in Ref.~\cite{Bini:2026kwz} indicates that, the model systematics observed in GW231123 can be reproduced in a controlled zero-noise simulation.  
The authors injected a signal generated with {\tt NRSur7dq4} model, and perform Bayesian inference with different waveform models.
The maximum likelihood waveforms
are very similar, despite of the large differences in the maximum likelihood parameters \cite{Bini:2026kwz}.
However, Ref.~\cite{Bini:2026kwz} mainly computed the waveform differences in time domain, without addressing whether such differences are measurable in realistic detector noise. 
This question is central to our residual test analysis: are these differences too small to leave a statistically significant residual? 
Motivated by this point, we first compare the whitened waveform differences between models and then perform injection tests to quantify the detectability of these differences. 

The waveform difference between different models, i.e., $\Delta_{\rm M} h$, is defined as
\begin{align}
    \Delta_{\rm M} h =  h^{\rm M_1}(t) -h^{\rm M_2}(t).
\end{align}
We follow Ref.~\cite{Bini:2026kwz} to take the best-fit waveform from the {\tt NRSur7dq4} model as the reference waveform, denoted by $h^{\rm M_1}$, which plays the role of the ``true signal" in our waveform difference comparison. 
We then compute the difference between this ``true signal" and the ``template", which is the best-fit waveform from the other models. 
In other words, the waveform difference here is treated as a signal-template difference in the following simulation tests.
It is an important approximation in our simulations.

Unlike Ref.~\cite{Bini:2026kwz}, where the best-fit waveforms were obtained by performing Bayesian inference on a {\tt NRSur7dq4} waveform injection, our comparison uses the posterior samples of the real GW231123 event. 
Thus, the signal-template differences studied here are directly inferred from the observed data. 
Notice that the difference between the {\tt NRSur7dq4} and {\tt IMRPhenomXO4a} waveforms here is comparable in size to that reported in Ref.~\cite{Bini:2026kwz} (see Fig.~1 in Ref.~\cite{Bini:2026kwz}), indicating that the strain-level model discrepancy in the real event is similar to that found in their injection analysis.
From this perspective, our approximation is realistic.

\begin{figure}
    \centering
    \includegraphics[width=\linewidth]{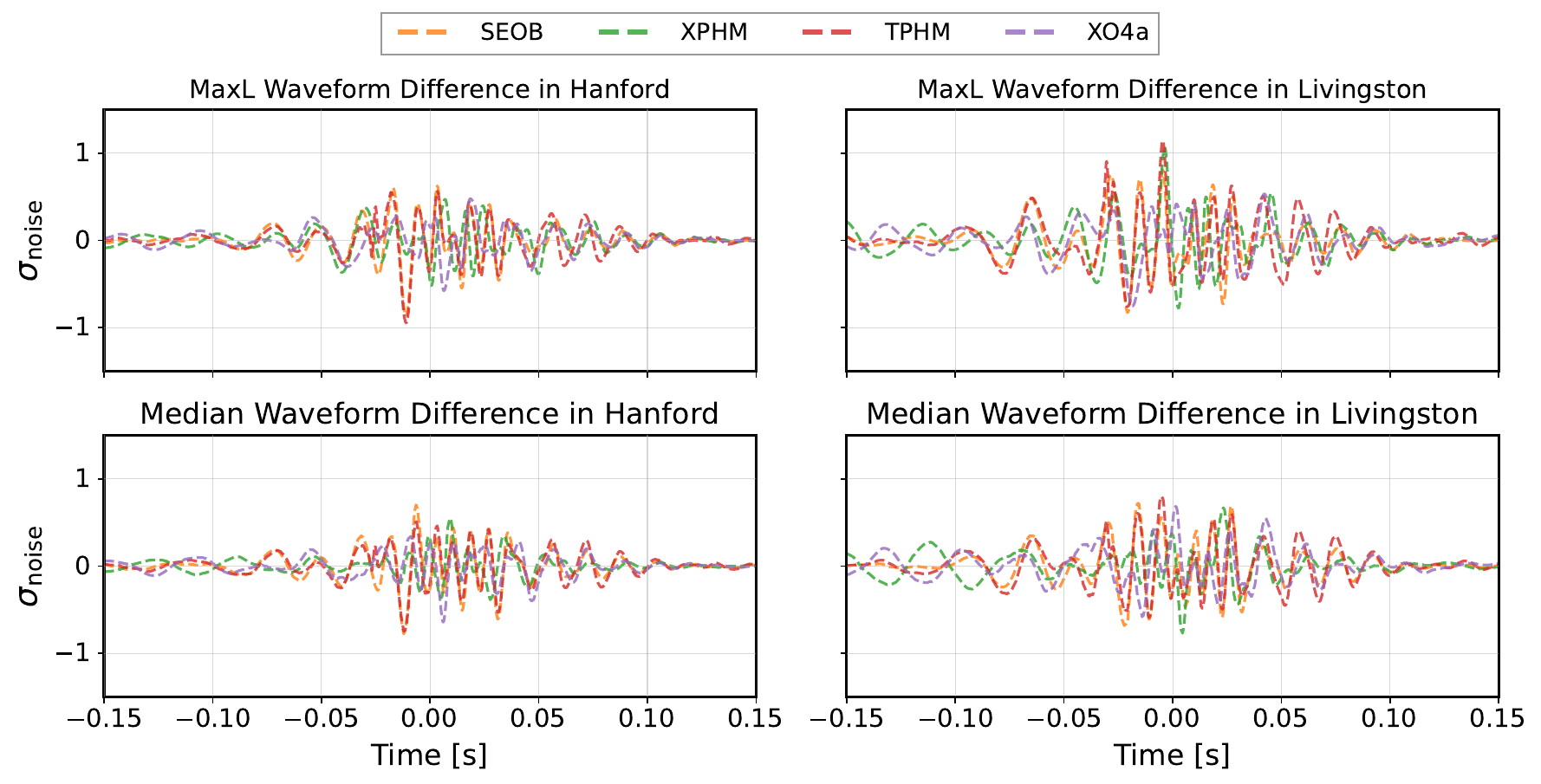} 
    \caption{ 
    Cross-model waveform differences for GW231123 in Hanford and Livingston. 
	We use {\tt NRSur7dq4} as the reference model and subtract the whitened waveforms of the other four models from it. 
    The upper panels show the differences computed using the maximum likelihood waveforms, while the lower panels show the differences computed using the median waveforms from the 100 highest likelihood samples. 
    The vertical axis is normalized by the whitened noise standard deviation. 
    Time is measured relative to 13:54:30.619 UTC, November 23rd 2023.
     }
    \label{fig:waveform_diff}
\end{figure}

The results are shown in the upper panels of Fig.~\ref{fig:waveform_diff}.
For Hanford, the difference between {\tt NRSur7dq4} model and {\tt IMRPHENOMXPHM } or {\tt IMRPHENOMXO4a} is $\lesssim 0.5 \sigma_{\rm noise}$. 
While the difference between {\tt NRSur7dq4} model and the other two models can reach approximately $1 \sigma_{\rm noise}$.
Overall, the waveform difference is relatively smaller in Hanford than that in Livingston.
We think it is due to the fact that Livingston has a higher sensitivity at the merger time of GW231123.

For comparison, we also show the difference between the median of the 100 highest likelihood waveforms after whitening in the lower panels of Fig.~\ref{fig:waveform_diff}.
Both the maximum likelihood and median waveform differences are concentrated near the merger and ringdown, where GW231123 has most of its observable power. 
Their amplitudes are typically below or comparable to the whitened noise level, showing that the different models can yield similar strain-level reconstructions. 
These cross-model differences are then treated as signal-template differences, thus are used to generate residuals in the injection tests below, allowing us to quantify how strongly such systematics would need to be amplified before becoming detectable by our residual tests.

We also compute the overlaps between the ``signal", i.e., best-fit waveform from the {\tt NRSur7dq4} model and the ``template", i.e., best-fit waveforms from other models. 
The results are summarized in Table~\ref{tab:overlap}.
The overlap values are all high, ranging from 0.963 to 0.987, indicating that the different models produce very similar strain-level waveforms.
The frequency overlap domain results are consistent with the time domain waveform differences shown in Fig.~\ref{fig:waveform_diff}.

\begin{table}[!htbp]
  \vspace{20pt}
  \centering
  \begin{tabular}{>{\centering\arraybackslash}p{1.8cm}
  >{\centering\arraybackslash}p{2.8cm}
  >{\centering\arraybackslash}p{2.8cm}
  >{\centering\arraybackslash}p{2.8cm}
  >{\centering\arraybackslash}p{2.8cm}
  }
  \hline
  detector & {\tt SEOBNRv5PHM} & {\tt IMRPHENOMXPHM }  & {\tt IMRPHENOMTPHM }   & {\tt IMRPHENOMXO4a}  \\
  \hline
  Hanford & 0.968 & 0.982  &   0.967   & 0.984  \\
  Livingston & 0.972 & 0.978  & 0.963  & 0.978 \\
  \hline
  \end{tabular}
    \caption{ Overlap between the maximum likelihood waveform in {\tt NRSur7dq4} model and those in other models. }  
  \label{tab:overlap}
\end{table}

To perform the injection test, we select 200 noise samples, $n_i(t)$, randomly around the event GW231123 in each detector, and these noise samples do not contain glitches.
The simulated residuals are computed by
\begin{align}
    \mathcal{R}_i = \alpha \Delta_{\rm M} h + n_i, \quad i=1,2,\cdot\cdot\cdot,200.
\end{align}
When the scaling factor $\alpha =0$, the residual reduces to noise only.
When $\alpha>1$, it corresponds to amplifying the signal-template difference artificially.
If the signal-to-noise ratio (SNR) of a GW231123-like source increases by a factor $\alpha$, because of closer distance or improved detector sensitivity, the difference would not generally be $\alpha \Delta_{\rm M} h$.
At higher SNR, the best-fit parameters of each model would shift, and part of the discrepancy could be absorbed by this parameter re-optimization.
Therefore, on one hand, the actual optimized difference is expected to be smaller than the frozen rescaling $\alpha \Delta_{\rm M} h$.
While, on the other hand, this waveform modeling systematics should be more measurable for larger SNR events.
In this sense, we expect the difference should be larger than $\Delta_{\rm M} h$ in general.
Our injection study, which uses $\alpha \Delta_{\rm M} h$, can be regarded as an upper-bound estimation of the detectability of such discrepancy.

\begin{figure}
    \centering
    \includegraphics[width=\linewidth]{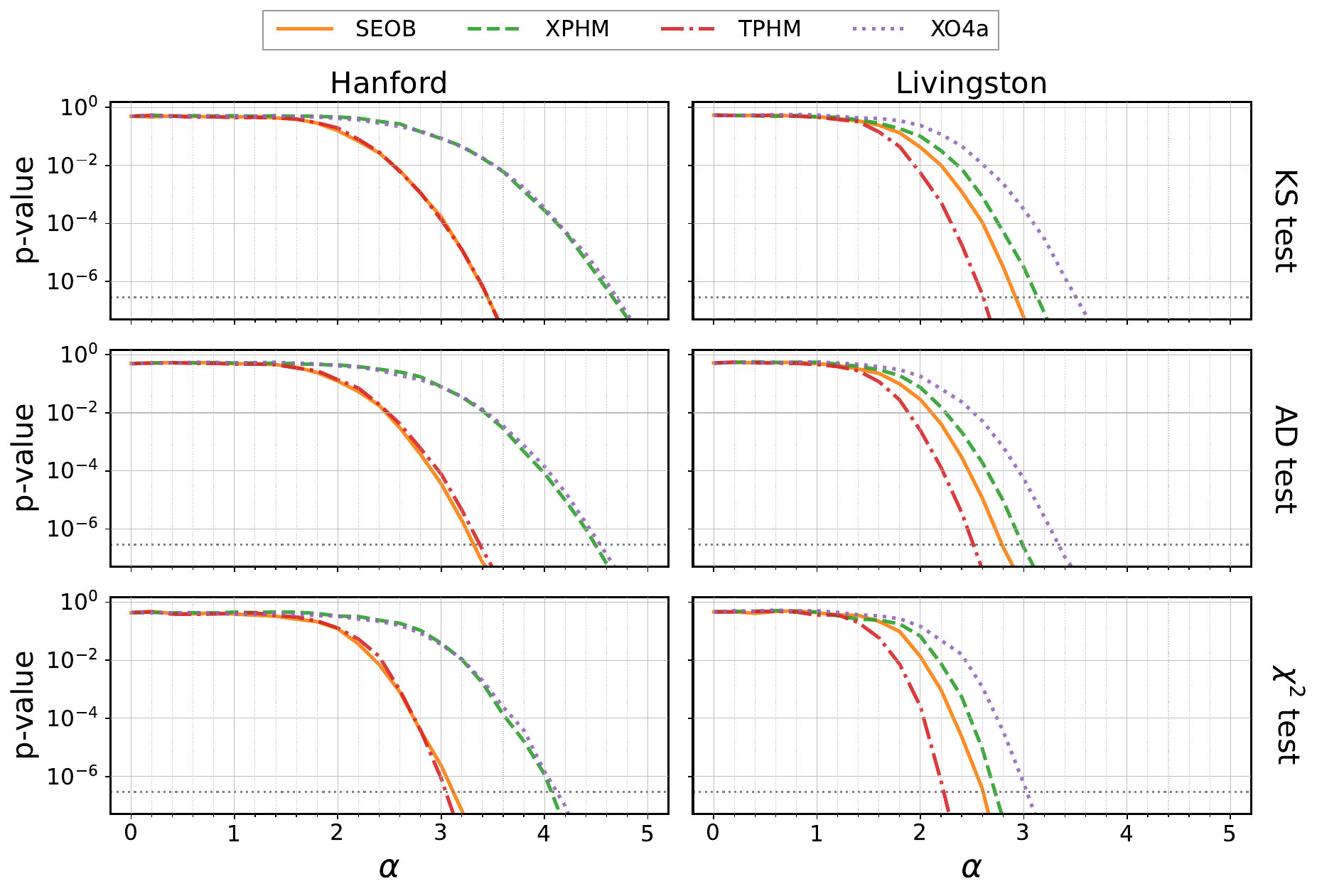} 
    \caption{ Median p-values of the injection tests as a function of the waveform-difference scale factor $\alpha$ in Hanford and Livingston. 
    For each model M, we inject scaled waveform differences, $\alpha \Delta_{\rm M} h$, into nearby noise samples and apply the KS, AD, and $\chi^2$ residual tests. 
    The three rows correspond to the three tests. 
    The curves show the median p-values over the 200 noise samples. 
    The horizontal gray dotted lines indicate the p-value threshold of $2.87\times10^{-7}$, corresponding to a $5\sigma$ significance level.
    }
    \label{fig:injection}
\end{figure}

In practice, we let the parameter $\alpha$ vary over the range from $0$ to $5$ with an interval (step size) of $0.2$.
After injecting into the 200 noise samples, the median values of the p-value of the residual tests for $\mathcal{R}_i$ with different $\alpha$, are shown in Fig.~\ref{fig:injection}.

As $\alpha$ increases, the simulated residual, $\mathcal{R}_i$, deviates more from the noise-only sample $n_i$.
Thus, as expected, the p-value curves decrease monotonically for all scenarios with the increase of $\alpha$.
A smaller waveform difference remains harder to distinguish from noise and therefore leads to a slower falloff of the p-value curve.
This behavior is clearly visible in the curves associated with {\tt IMRPhenomXO4a}. 
Since the waveform difference between the best-fit {\tt NRSur7dq4} waveform and the {\tt IMRPhenomXO4a} waveform is the smallest among the models considered here, the corresponding p-value curves decrease most slowly. 
Conversely, waveform models with larger differences from {\tt NRSur7dq4} reach small p-values at smaller values of $\alpha$. 
In Hanford, the waveform differences for {\tt SEOBNRv5PHM} and {\tt IMRPhenomTPHM} are very similar, and their p-value curves nearly overlap, consistent with the time-domain waveform differences shown in Fig.~\ref{fig:waveform_diff}.
We also find similar behavior for {\tt IMRPHENOMXPHM} and {\tt IMRPhenomXO4a} in Hanford.

The horizontal gray dotted lines in each subplots indicate a p-value of $2.87\times10^{-7}$, corresponding to a significance level of $5\sigma$.
Once the p-value drops below this line, the injected signal-template difference becomes statistically distinguishable from noise by the corresponding residual test. 
The figure shows that, for most models and tests, the waveform-difference residual must be amplified by a factor of roughly a few before it becomes significant. 
Equivalently speaking, the current signal-template differences in GW231123 are generally too small to be detected directly by the residual tests, but a GW231123-like event with a larger effective SNR could make such differences measurable.

\section{Conclusions and Discussions}
\label{sec:disc}

As an exceptional compact binary coalescence event, the source of GW231123 has unusually large component masses and high spins, placing it in a challenging region of parameter space for waveform modeling.
As shown in Fig.~\ref{fig:para}, applying five different waveform models to infer properties of GW231123 leads to significantly different posterior distributions for the source parameters.
This raises one important question: whether the reconstructed waveforms in these models provide sufficiently accurate descriptions of the observed signal. 
It motivates our residual analysis of GW231123.

Previous residual tests typically subtract only the maximum likelihood waveform from the strain data and then examine whether the remaining residual is consistent with noise. 
In this work, we extend this procedure by subtracting the 100 highest likelihood waveforms for each model, rather than relying on a single maximum likelihood waveform. 
Given the extreme properties of GW231123, the local high likelihood waveform manifold may be broad or irregular, and the numerically identified maximum likelihood sample may not be uniquely representative of the signal reconstruction. 
It is therefore important to test an ensemble of high-likelihood waveforms, which makes the residual analysis more robust against sampling fluctuations and waveform reconstruction uncertainty. In Fig.~\ref{fig:waveforms}, we show both the maximum likelihood waveform and the median waveform constructed from the 100 highest likelihood samples, while Fig.~\ref{fig:overlap} presents their frequency-domain overlaps within each model. 
Both the maximum likelihood and median waveforms reproduce the main signal cycles well in the time domain, and the high overlap values indicate that the 100 highest likelihood waveforms are internally consistent within each waveform model.

After subtracting the 100 highest likelihood waveforms, we apply the q-transform to the corresponding 100 residuals for each detector and each waveform model. 
The median and pixel-wise standard deviation of the normalized q-transform energies are shown in Figs.~\ref{fig:q_median} and \ref{fig:q_std}, respectively. 
The median maps show the typical residual structure after waveform subtraction, while the standard-deviation maps quantify the internal variation of the waveforms within each model in the time-frequency plane.
We further show the distributions of the normalized q-transform energies in Fig.~\ref{fig:q_hist}, providing a visual comparison with the exponential distribution expected for stationary white noise. 
To quantify this consistency, we apply three goodness-of-fit tests, i.e., the KS, AD, and $\chi^2$ tests. 
The resulting p-value distributions are shown in Fig.~\ref{fig:p_res}, and representative p-values are summarized in Fig.~\ref{fig:p_res_m}. The p-values are generally large, especially for the median residuals, indicating that the high-likelihood waveforms provide good descriptions of the observed signal at the current detector sensitivity. 
This analysis therefore serves not only as a residual consistency test, but also as a test of the internal stability of the high-likelihood waveform reconstructions within each model.

Next, we examine the waveform discrepancies across different waveform models. Following Ref.~\cite{Bini:2026kwz}, we take the best-fit waveform from the {\tt NRSur7dq4} model as the reference waveform, which plays the role of the “true signal” in our comparison, and regard the best-fit waveforms from the other models as alternative waveform templates. 
Unlike Ref.~\cite{Bini:2026kwz}, where the comparison was performed using a full Bayesian inference on {\tt NRSur7dq4} injection, our waveform differences are computed from the posterior samples of the real GW231123 event. 
Nevertheless, the resulting waveform differences are comparable to those reported in Ref.~\cite{Bini:2026kwz}  (see Fig.~\ref{fig:waveform_diff} in this work and Fig.~1 of Ref.~\cite{Bini:2026kwz}). 
To quantify whether such waveform discrepancies are measurable in realistic detector noise, we construct simulated residuals by injecting the scaled waveform difference $\alpha \Delta_{\rm M}h$ into nearby noise samples. 
We then apply the KS, AD, and $\chi^2$ tests to these simulated residuals for different values of scaling factor $\alpha$, and show the median p-values over the 200 noise samples in Fig.~\ref{fig:injection}. 
Although the scaling factor artificially amplifies the waveform difference, it has a useful physical interpretation: for a GW231123-like source, waveform-model systematics would become more visible if the source were closer or if the detector sensitivity were improved. 
In a true higher-SNR observation, part of the waveform discrepancy could be absorbed by re-optimizing the best-fit parameters, so $\alpha \Delta_{\rm M}h$ should be regarded as an upper-bound estimate of the detectable signal-template difference. 
Our injection tests show that amplifying the present waveform discrepancies by a factor of a few is sufficient to produce a $5\sigma$-level deviation of the residuals from the noise expectation.

Last but not least, it is important to emphasize what residual tests can and cannot tell us. 
A residual test, including the ensemble-based extension introduced here, is a powerful tool for checking the consistency between a given waveform model and the observed strain data — in other words, whether the residual is statistically compatible with detector noise. 
However, it is not a model-selection or parameter-estimation tool. 
When multiple waveform models all pass the residual test, as we find for GW231123, this does not mean that their inferred source parameters are equally reliable, nor does it indicate which model best describes the true signal. 
It simply reflects the fact that the strain-level differences between the high likelihood waveforms of different models are too small to be detected at the current sensitivity.

Note that, this work involves a large-scale implementation of the residual-test framework. 
We perform residual tests for the 100 highest likelihood waveforms in each detector and for all five waveform models. 
In addition, we carry out injection tests for the four cross-model waveform discrepancies, using 200 nearby noise samples for each of the 26 discrete values of the scale factor $\alpha$. 
The feasibility of these extensive calculations demonstrates a major advantage of our residual-test method: it is fast, computationally inexpensive, and therefore well suited for ensemble-based robustness studies and injection-based sensitivity tests.

Looking ahead, this framework can be extended in several directions. 
First, the ensemble-based residual test can be applied systematically to future high-SNR, high-mass, or high-spin events, for which waveform systematics are expected to become increasingly important. 
Second, the injection strategy developed here can be used to forecast the detectability of waveform discrepancies with improved detector sensitivities, providing a practical way to assess when current waveform models will no longer be sufficient. However, in our injection tests we approximate the cross-model best-fit waveform difference as a proxy for the signal-template discrepancy. 
This approximation should be further validated with full Bayesian inference, in which simulated signals are reanalyzed using different waveform models. 
Such studies would also allow one to verify the p-value evolution with the scaling factor under a fully re-optimized parameter estimation procedure. 
Finally, the method can be combined with more advanced waveform families, including eccentric, lensing-motivated, and modified-gravity waveform models. 
It can therefore serve as a flexible diagnostic tool for testing waveform accuracy and identifying possible unmodeled physics in future GW observations.

\section*{Acknowledgments}

D. Liang thanks Prof. Alan J. Weinstein for the valuable discussions, which inspired this work.
This work is supported by the National Natural Science Foundation of China (12405065, 12305066, 12465013).
H.-T. Wang is supported by ``the Natural Science Foundation of Liaoning Province" (Grant No. ZX20250217) and ``the Fundamental Research Funds for the Central Universities" at Dalian University of Technology.
J. Qin and Z.-F. Mai are supported by the ``Hanji" Action Plan (Guangxi Basic Research Program, Grant No. 2026GXNSFBA00640240), the Guangxi Science and Technology Innovation Platform Program (Leitai Action Plan, Grant No. Guike LT2600640026) and the ``Guangxi Highland of Innovation Talents" Program.


%

\end{document}